\documentclass[amsmath,amssymb,aps]
{revtex4-2}

\usepackage{graphicx}
\usepackage{dcolumn}
\usepackage{bm}
\usepackage{chemformula}
\usepackage{multirow}

\usepackage[lofdepth,lotdepth]{subfig}

\begin{document}

\title{Ab-initio study of the effect of bromide mixing into \ch{RbPbI_3} on the structural, electronic and optical properties}%

\author{Anupriya Nyayban}%
\email{$\rm{anupriya_rs@phy.nits.ac.in}$}
\author{Subhasis Panda}
\email{subhasis@phy.nits.ac.in}
\affiliation{Department of Physics\\National Institute of Technology Silchar, Assam, 788010, India }
\author{Avijit Chowdhury}
\email{avijitiacs@gmail.com}
\affiliation{Department of Physics\\National Institute of Technology Silchar, Assam, 788010, India \\ and\\Department of Condensed Matter Physics and Material Sciences, S.N. Bose National Centre for Basic Sciences, JD Block, Sector III, Salt Lake City,
Kolkata 700106, India}
\date{\today}%


\begin{abstract}
The ultra-high efficiency and cost-effective photovoltaics based on halide preovskites have brought a revolution to ongoing photovoltaic research, surpassing the expectations of the scientific community. However, structural stability is a severe issue that hinders their wide-scale integration at the device level. Compositional engineering with the halide mixing has become an efficient way to deal with this issue without compromising device efficiency. Herein, the structural, electronic and optical properties of the bromide mixed orthorhombic \ch{RbPb(I_{1-x}Br_x)_3} (where, $\ch{x}=0.25$, $0.50$ and $0.75$) are calculated using the density functional theory. The electronic bandstructure and density of states (DOS) are calculated using both PBE (Perdew-Burke-Ernzerhof) and TB-mBJ (Tran Blaha modified Becke Johnson) potential. The lowest energy bandgaps of $2.288$ and $2.986$ eV for bromide mixing of $\ch{x}=0.50$ are obtained using PBE and TB-mBJ, respectively. In contrast, the mixed bromide phases possess a smaller effective mass, facilitating a better carrier transport through the mixed halide. Using PBE, the excitons appear to be the Mott-Wannier type. However, the TB-mBJ predicts the exciton to be Frenkel type for bromide mixing of $\ch{x}=0.75$ and a Mott-Wannier type for all other mixing under study. The spectroscopic limited maximum efficiency (SLME) is observed to be at the highest values of $14.0$\% and $4.1$\% for the equal admixture of \ch{I} and \ch{Br} using PBE and TB-mBJ, respectively. The calculated properties are consistent with the reported data of the similar structures. 
\end{abstract}

\maketitle

\section{Introduction}
Hybrid halide perovskites have been the most studied material recently in the photovoltaic group due to its superior properties for solar cell applications such as suitable bandgap and band alignment \cite{ref1,ref2,ref3,ref4}, strong absorption \cite{ref5,ref6}, better charge transport \cite{ref7,ref8,ref9,ref10,ref11} and lower defect properties \cite{ref12,ref13,ref14}. Moreover, halide perovskites are processed by solution methods with the lower cost and compositional materials are abundant in nature \cite{ref15,ref16}. In spite of these excellent properties, the instability of organic-inorganic halide perovskites due to heat, oxygen, moisture and light \cite{ref17,ref18,ref19,ref20} prevent them to be commercialized. Normally, inorganic materials exhibit better stability in terms of temperature than the organic one \cite{ref21,ref22,ref23,ref24}. The unencapsulated \ch{MAPbBr_3} based device exhibit a faster $55$\% decay of the phototcurrent density compared to the maximum value during an illumination of $5$ h whereas \ch{CsPbBr_3} based one shows only $13$\% decay \cite{ref25}.
Additionally, \ch{MAPbBr_3} based cell faces $85$\% efficiency loss  in an environment of $60-70$ \% relative humidity for 2 weeks whereas \ch{CsPbBr_3} based cell shows almost no loss \cite{ref25}. However, all inorganic perovskites were not popular due to their lower efficiency. Recently, the \ch{Cs}-based perovskite is reported \cite{ref49} in 2016 to achieve the efficiency of $10$\% by tuning the bandgap with adjusting ratio of iodide-bromide. The mixed halide perovskite based solar cells which is deposited using all-vacuum-deposition technique, also shows the surplus efficiency of $11$\% \cite{ref50}. In the similar time, \ch{CsPbI_3}-based quantum dot solar cell not only exhibits efficiency of $11$\% but also it poses the phase stability \cite{ref51}. Thereafter, these research efforts have inspired researchers to have an in-depth insight into the fundamental properties of all inorganic perovskites for better functionality. \\ The halide mixing i.e. the mixing of multiple halogen elements are very popular in recent trend due to the several beneficial effects. Firstly, the halide mixing increases the stability e.g. \ch{MAPbI_{3-x}Cl_x} are reported \cite{ref26} to posses remarkable stability in air as compared to \ch{MAPbI_3}. \ch{MAPbI_3} with $20-29$\% \ch{Br} mixing is also found \cite{ref27} to improve the stability tremendously without changing the efficiency. Secondly, mixed halide perovskites enhance the transport of charge carriers. The longer charge carrier diffusion length is observed \cite{ref28,ref29,ref30} in \ch{MAPbI_{3-x}Cl_x} than \ch{MAPbI_3}. The improved mobility and the reduced recombination rates are also reported \cite{ref31,ref32} in both \ch{MAPbI_{3-x}Br_x} and \ch{MAPbBr_{3-x}Cl_x}. Third, bandgap can be tuned by halide mixing in halide perovskites. Chloride mixed \ch{MAPbBr_3} i.e. \ch{MAPbBr_{3-x}Cl_x} are observed \cite{ref32} to exhibit better photovoltaic performance having the higher open circuit voltage of $1.5$ eV and the reduced short circuit current. \ch{MAPbI_{3-x}Br_x} are also observed \cite{ref33} to be suitable for the colorful solar cell design in order to found the energy saving buildings. These reported findings suggest that the halide mixing into the inorganic halide perovskites could be an efficient way to improve the photovoltaic  performances. Moreover, no literature is found for the iodide-bromide mixed \ch{RbPbI_3} i.e. \ch{RbPb(I_{1-x}Br_x)_3} towards the electronic and optical properties.\\In this work, we have systematically studied the effect of the bromide mixing into \ch{RbPbI_3} (with \ch{Br} concentration of $0.25$, $0.50$ and $0.75$) on the structural, electronic and optical properties. The work is arranged in the following manner: Section II describes the computational methods; Section III-A provides the structural information; Section III-B describes the electronic structures in detail including the density of states (DOS) and the bandstructures; Section III-C informs extensively regarding the optical properties, effective masses of the photo-generated charge carriers, the spectroscopic limited maximum efficiency (SLME); and at last the work is summarized and concluded in the section IV. 
\section{Computational details}
At first, $2 \times 2 \times 2$ supercell is constructed for \ch{RbPb(I_{1-x}Br_x)_3} (when $\ch{x}=0.25$, $0.50$ and $0.75$) from an optimized orthorhombic \ch{RbPbI_3} of $Pnma$ space group \cite{ref34}. Then the supercell is relaxed and optimized. All the density functional theory (DFT) based calculations have been performed using WIEN2k \cite{ref35}. A muffin tin radius of $2.50$ {\AA} is set for all the atoms whereas a $\rm{Rk_{max}}$ of $7$ is set for all the structures. Perdew-Burke-Ernzerhof (PBE) \cite{ref36} exchange correlation functional has been reported \cite{ref37,ref38,ref39,ref40} to estimate the bandgap accurately for the halide perovskites without considering the spin orbit coupling (SOC) effect. Again PBE including SOC are reported \cite{ref41} to underestimate the bandgap but does not alter the pattern of the bandstructure. Additionally the hybrid functional is also reported \cite{ref42} to overestimate the bandgap whereas less computationally demanding TB-mBJ (Tran Blaha modified Becke Johnson) \cite{ref43} potential is found to estimate the bandgap more accurately \cite{ref44,ref45}. Therefore, PBE and TB-mBJ potentials are used to find both the electronic and optical properties for all \ch{I}-\ch{Br} mixed structures. The electronic properties are estimated with a kmesh of $1 \times 5 \times 11$ for $\ch{x}=0.50$ and $2 \times 3 \times 14$ for both $\ch{x}=0.25$ and $0.75$, respectively. For the calculation of optical properties, a denser kmesh of $3 \times 11 \times 25$ for $\ch{x}=0.50$ and $4 \times 7 \times 31$ for other two structures, respectively. 
\section{Results and observations}

\subsection{Optimized Structures}

\ch{RbPbI_3} have orthorhombic \ch{NH_4CdCl_3} type structure of $Pnma$ space group at room temperature. A supercell of $2 \times 2 \times 2$ has been considered from the optimized pristine \ch{RbPbI_3} unit cell \cite{ref34}. The iodide has been gradually replaced with bromide by the concentration \ch{x} of \ch{Br}. PBE-GGA have been used to find the structural properties. The total energy with the variation of the supercell-volume are fitted using Birch-Murnaghan equation of state \cite{ref46} as shown in FIG. S1 in the supporting information. The optimized supercells are also depicted in FIG. S2 in the supporting information. The estimated lattice parameters, volume of supercell ($V_0$), bulk modulus ($B$) and pressure derivative ($B^{'}$) for \ch{RbPb(I_{1-x}Br_x)_3} are listed in TABLE \ref{tab:latibr}. The lattice parameters decrease with the increasing \ch{Br}-concentrations except for $\ch{x}=0.50$. The decrease in lattice parameters is attributed to the reduced \ch{Br} radii.

\begin{table}[h!]
\caption{Lattice parameters for \ch{RbPb(I_{1-x}Br_{x})_3} systems} 
\label{tab:latibr}
\begin{tabular}{ccccccccc}
\hline
\hline
\ch{x} & a (\AA) & b (\AA) & c (\AA) & $V_{0}$ (${\AA}^3$) & $B$ (GPa) & $B^{'}$  (GPa)\\
\hline
$0.25$ & $34.060$ & $20.149$ & $4.681$ & $3212.151$ & $12.8517$ & $7.5097$\\
$0.50$ & $34.533$ & $10.214$ & $4.746$ & $1673.957$ &  $11.8370$ & $10.1306$\\
$0.75$ & $33.789$ & $19.989$ & $4.643$ & $3136.202$ & $11.7332$ & $7.2820$\\
\hline
\hline
\end{tabular}
\end{table}

\subsection{Electronic Properties}
\begin{figure}[h!]
\centering
  \subfloat[\ch{RbPb(I_{0.75}Br_{0.25})_3}.]{\includegraphics[scale=0.5]{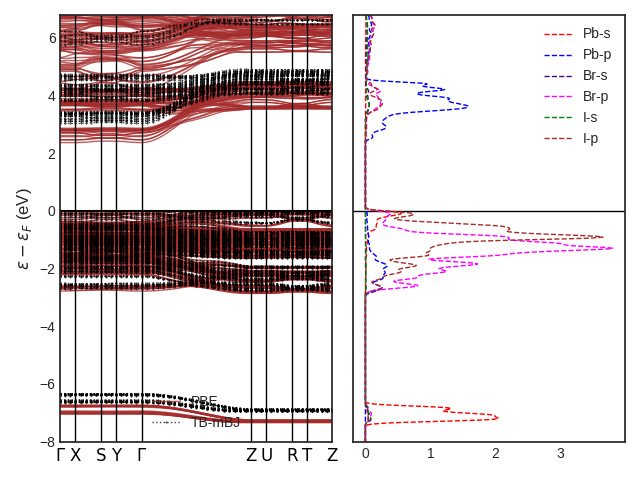}\label{fig:bdbr25}}
  \subfloat[\ch{RbPb(I_{0.50}Br_{0.50})_3}.]{\includegraphics[scale=0.5]{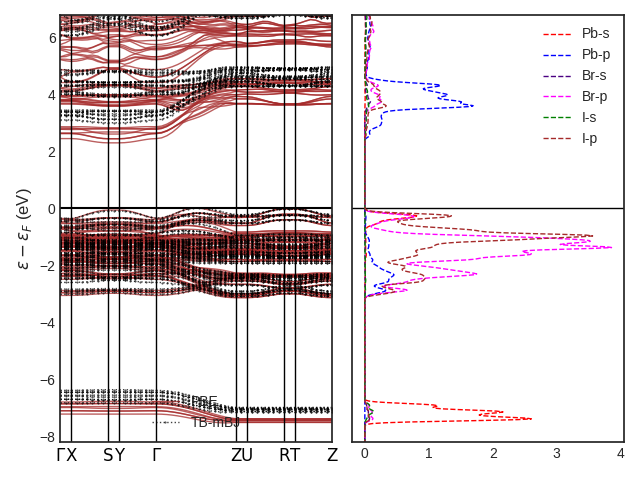}\label{fig:bdbr50}}
    \hspace{0.1cm}
  \subfloat[\ch{RbPb(I_{0.25}Br_{0.75})_3}.]{\includegraphics[scale=0.5]{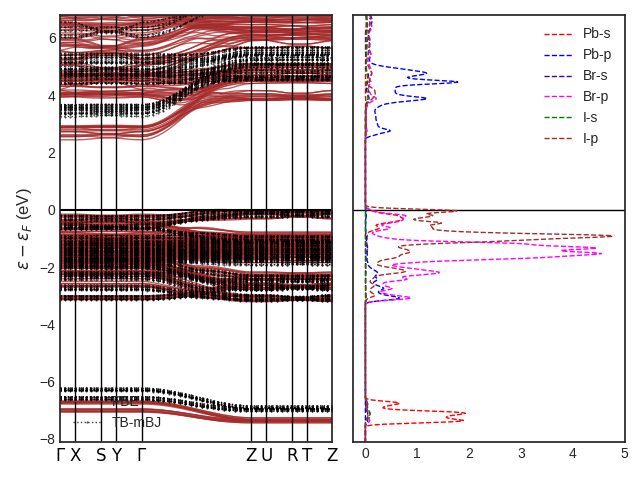}\label{fig:bdbr75}}    
    \subfloat[\ch{RbPb(I_{1-x}Br_{x})_3}.]{\includegraphics[scale=0.5]{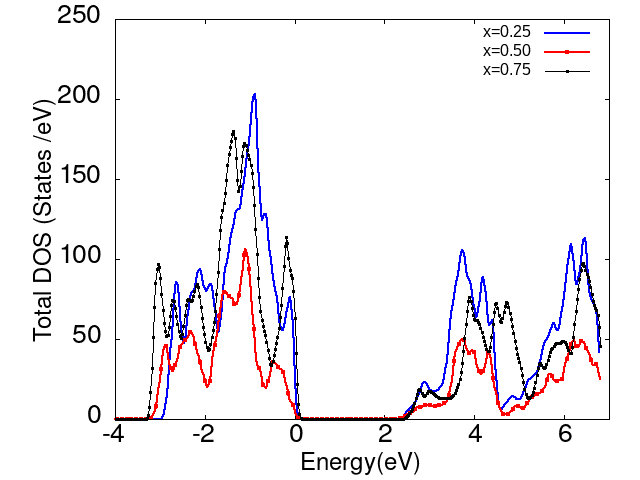}\label{fig:tdibr}} 
   \caption{The band structures along with the partial density of states (PDOS) and total density of states (TDOS) for all the bromide mixed systems.}\label{fig:bdibr}
\end{figure}

The electronic properties are investigated with both PBE and TB-mBJ potentials. The electronic band structures along the high symmetry $k$-point $\Gamma \rightarrow X \rightarrow S \rightarrow Y \rightarrow \Gamma \rightarrow Z \rightarrow U \rightarrow R \rightarrow T \rightarrow Z$ are shown in FIG. \ref{fig:bdibr}. The band gap estimated with PBE is ranging from $2.288$ to $2.435$ eV while that with TB-mBJ potentials is from $2.986$ to $3.253$ eV for the increasing \ch{Br}-content. The valence band maximum (VBM) is observed at a $k$ point between $\Gamma \rightarrow Z$ path for all the halide mixed systems under consideration. Again, the conduction band minimum (CBM) is found at $\Gamma$ and $S$ for ($x=0.25$ \& $0.75$) and ($\ch{x}=0.50$), respectively, confirming all \ch{I}-\ch{Br} mixed structures to have indirect band gap. All the calculated band gap values are listed in TABLE \ref{tab:bgpibr}. The band gap decreases with the increase of the percentage of \ch{Br} due to the reduced lattice parameters. The lowest band gap is observed for the equal mixture of \ch{I} and \ch{Br}. The band gap estimated with TB-mBJ potential agrees well with the experimental values for the halide mixed \ch{CsPbI_3} systems \cite{ref47}. TB-mBJ potential uplifts the CBM towards the higher energy and thus, widens the band gap values. No significant changes in the valence band as well as in the nature of the bandstructure is observed. The band gap of the \ch{RbPb(I_{1-x}Br_x)_3} is reduced compared to that of both pristine \ch{RbPbI_3} and \ch{RbPbBr_3} \cite{ref34}.  The similar behavior of the bandstructures calculated with both PBE and TB-mBJ is also observed for \ch{RbPbI_3} \cite{ref34}. 
 
\begin{table}[h!]
\caption{The band gap $E_g$ in eV for all \ch{RbPb(I_{1-x}Br_x)_3} mixed systems}
\label{tab:bgpibr}
\begin{tabular}{cccc}
\hline
\hline
 & \ch{x}=0.25 & \ch{x}=0.50 & \ch{x}=0.75\\
\hline
PBE & $2.366$ & $2.288$ & $2.435$\\
TB-mBJ & $3.025$ & $2.986$ & $3.253$\\
\hline
\hline
\end{tabular}
\end{table}

The atomic orbital contribution has been explored using the density of state (DOS) estimation  in order to investigate the electronic structure in brief. Partial density of states (PDOS) calculated with PBE and bandstructures calculated using PBE and TB-mBJ for all the mixed halide systems are shown in FIG. \ref{fig:bdibr}. \ch{Rb} has no contribution in either VBM or CBM and hence, the orbital contributions of \ch{Pb}, \ch{I} and \ch{Br} are only shown. The first CB is mainly dominated with $\rm{6p}$ orbital of \ch{Pb}. The  range of the first CB are from $2.48$ to $4.47$, $2.52$ to $4.61$ and $2.53$ to $5.10$ eV corresponding to the increasing \ch{Br}-content of $0.25$, $0.50$ and $0.75$, respectively. The first valence band (VB) is dominantly contributed by both \ch{Br}-$\rm{4p}$ and \ch{I}-$\rm{5p}$ with a small contribution of \ch{Pb}-$\rm{6p}$ towards the lower energy region of the first VB. \ch{Pb}-$\rm{6s}$ contributes to the second VB. The first VB expands from $0.00$ to $-2.82$, $-3.06$ and $-3.21$ eV for $\ch{x}=0.25$, $0.50$ and $0.75$, respectively. \ch{I}-$\rm{5p}$ and \ch{Br}-$\rm{4p}$ contributes more to the uppermost and lowermost region of the first VB, respectively. 

The total density of states (TDOS) for all \ch{RbPb(I_{1-x}Br_x)_3} systems are shown in FIG.  \ref{fig:tdibr}. The increase in TDOS is higher for both $\ch{x}=0.25$ and $0.75$ than that for $\ch{x}=0.50$. The VB rises higher than the CB. The higher and lower concentration of \ch{Br} exhibit the higher probability of larger carrier concentrations due to the larger TDOS and therefore, can increase the transport properties too. The similar phenomena are also observed for \ch{MAPb_{1-x}Ge_xI_3} \cite{ref48}.   

\subsection{Optical Properties}
\begin{figure}[h!]
\centering
  \subfloat[$\varepsilon_{2}$ estimated with PBE for \ch{RbPb(I_{1-x}Br_{x})_3}]{\includegraphics[width=6cm,height=7cm]{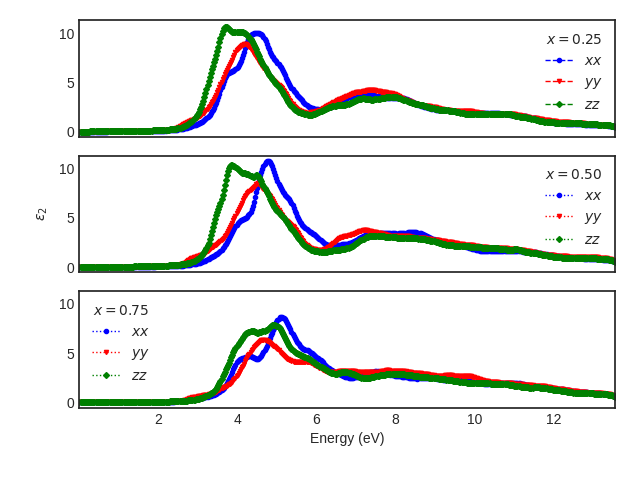}\label{fig:e2ibrp}}
  \subfloat[$\alpha$ calculated with PBE for \ch{RbPb(I_{1-x}Br_{x})_3}]{\includegraphics[width=6cm,height=7cm]{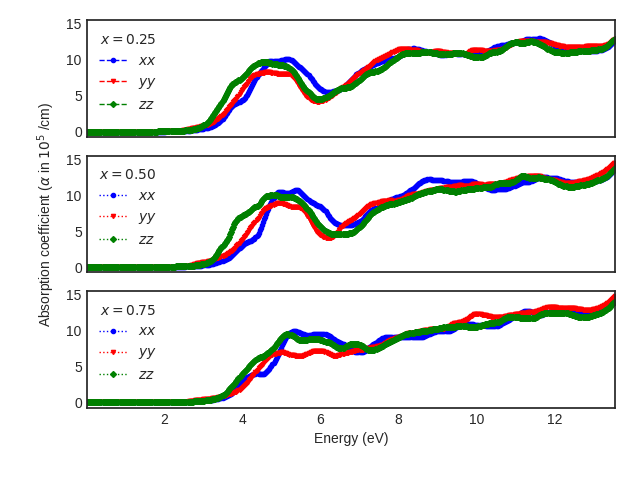}\label{fig:abibrp}}
  \hspace{0.1cm}
     \subfloat[$\varepsilon_{2}$ calculated with TB-mBJ for \ch{RbPb(I_{1-x}Br_{x})_3}]{\includegraphics[width=6cm,height=7cm]{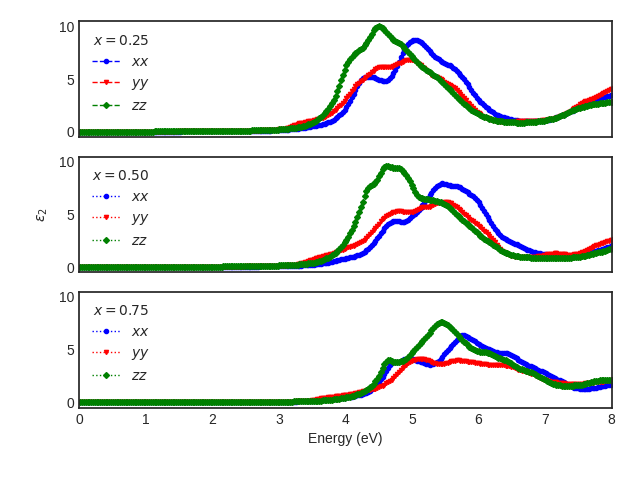}\label{fig:e2ibrm}}
      \subfloat[$\alpha$ estimated with TB-mBJ for \ch{RbPb(I_{1-x}Br_{x})_3}]{\includegraphics[width=6cm,height=7cm]{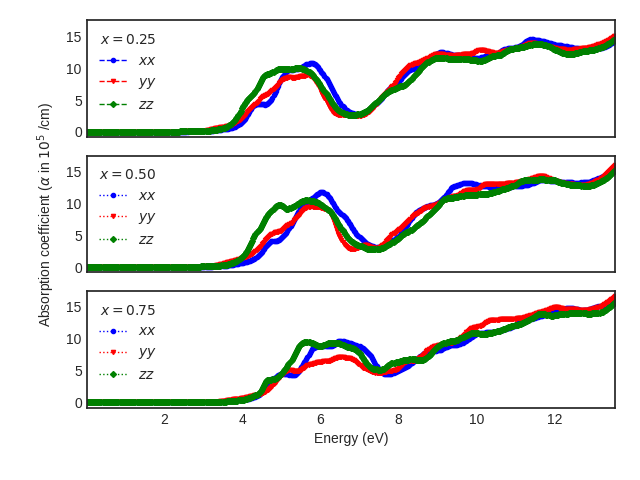}\label{fig:abibrm}}
   \caption{The variation of the imaginary part of the dielectric function ($\epsilon_{2}$) and the absorption coefficient ($\alpha$) with the photon energy along three different polarization direction for $\ch{x}=0.25$, $0.50$ and $0.75$, respectively.}\label{fig:e2abibr}
\end{figure}

The optical properties for all \ch{I}-\ch{Br} mixed \ch{RbPb(I_{1-x}Br_x)_3} systems are investigated with both PBE and TB-mBJ potential. All the optical properties are estimated using the complex dielectric function ($\epsilon(\omega)=\epsilon_{1}(\omega)+i\epsilon_{2}(\omega)$). At first, the momentum matrix using the occupied and unoccupied states are used to calculate the imaginary part of the dielectric function ($\epsilon_{2}(\omega)$). The variation of $\epsilon_{2}(\omega)$ with the photon energy are calculated using both PBE and TB-mBJ potential as show in FIG. \ref{fig:e2ibrp} and FIG. \ref{fig:e2ibrm}, respectively. The strongest peaks of $\epsilon_{2}(\omega)$ calculated with PBE are observed at $4.503$, $4.204$, $3.714$; $5.102$, $4.667$, $4.912$; $4.776$, $4.531$, $3.878$ eV for bromide concentrations of $0.25$, $0.50$ and $0.75$ along xx, yy and zz polarization directions, respectively. The spectrum of $\epsilon_{2}(\omega)$ calculated with TB-mBJ potential has the strongest peaks at the photon energy of $5.048$, $4.966$, $4.503$; $5.755$, $5.129$, $5.429$; $5.455$, $5.510$, $4.639$ eV along three different directions for the increasing \ch{Br} concentrations of $0.25$, $0.50$ and $0.75$, respectively. The trend of observed peaks in $\epsilon_{2}(\omega)$ are the result of the increasing bandgap with the increasing \ch{Br} percentage (except at $\ch{x}=0.50$).  

The real part of the dielectric function ($\epsilon_{1}(\omega)$) is calculated using Kramer-Kronig relation, once $\epsilon_{2}(\omega)$ has been estimated. Then extinction coefficient ($k(\omega)$), refractive index ($n(\omega)$), absorption coefficient ($\alpha(\omega)$) and reflectivity ($R(\omega)$) are calculated using the real ($\epsilon_1(\omega)$) and imaginary part ($\epsilon_2(\omega)$) of the dielectric function \cite{ref49}.

The spectrum of the absorption coefficient calculated using PBE and TB-mBJ are shown in FIG. \ref{fig:abibrp} and FIG. \ref{fig:abibrm}, respectively. The strongest peaks of the absorption spectrum calculated with PBE correspond to the photon energy of $5.157$, $4.667$, $4.558$; $5.401$, $4.966$, $4.803$; $5.320$, $5.891$, $5.157$ eV along xx, yy and zz directions for $\ch{x}=0.25$, $0.50$ and $0.75$, respectively. In case of the absorption spectrum calculated with TB-mBJ potential, the strongest peaks along three different directions are observed at $5.782$, $5.674$, $5.429$; $6.055$, $5.782$, $5.674$; $6.517$, $6.517$, $6.435$ eV for the increasing \ch{Br} percentage of $0.25$, $0.50$ and $0.75$, respectively. The transition of electrons from \ch{I}-5p, \ch{Br}-4p and \ch{Pb}-6s mixed valence state to   the conduction state consisting of \ch{Pb}-6p is mainly dominant for the strongest peaks of the absorption spectrum in all \ch{RbPb(I_{1-x}Br_x)_3} systems. The blueshift towards the higher energy region is observed for all \ch{I}-\ch{Br} mixed systems with the increasing \ch{Br} percentage  and this shift is attributed to the increasing bandgap. The absorption spectra shows that iodide-bromide mixed \ch{RbPb(I_{1-x}Br_{x})_3} systems can absorb photons from wider energy region (visible and ultraviolet) as compared to the pristine \ch{RbPbI_3} case. The nature of the absorption spectrum calculated with both PBE and TB-mBJ are similar whereas the blueshift toward the higher energy for TB-mBJ potential is more than that for PBE. 
Furthermore, the integration over the absorption coefficient along near-infrared ($0-1.7$ eV), visible ($1.7-3.3$ eV) and ultraviolet ($3.3-5.0$ eV) regions of the solar spectrum are calculated for all the considered systems and listed in TABLE \ref{tab:iiibr}. The average integrated intensity decreases with the increasing concentrations of \ch{Br}. The average absorption intensity decreases more with the bromide mixing in \ch{RbPbI_3} than that for the pristine \ch{RbPbI_3}. 

\begin{table}[h!]
\caption{The static dielectric constant, reflectivity, refractive index, binding energy of excitons and the radius of the lowest bound states for all the systems.}
\label{tab:e1rnibr}
\begin{tabular}{ccccccccccc}
\hline
\hline
& \ch{x} & \multicolumn{3}{c}{$\epsilon_{1}(0)$} & \multicolumn{3}{c}{$R(0)$} & \multicolumn{3}{c}{$n(0)$}\\
\hline
 & & $(100)$ & $(010)$ & $(001)$ & $(100)$ & $(010)$ & $(010)$ & $(100)$ & $(010)$ & $(010)$\\
 \hline
\multirow{ 2}{*}{$0.25$} & PBE & $4.673$ & $4.625$ & $5.107$ & $0.135$ & $0.133$ & $0.149$ & $2.162$ & $2.151$ & $2.260$\\ 
 & TB-mBJ & $3.818$ & $3.757$ & $4.085$ & $0.104$ & $0.102$ & $0.114$ & $1.954$ & $1.938$ & $2.021$\\ 
\multirow{ 2}{*}{$0.50$} & PBE & $4.302$ & $4.313$ & $4.814$ & $0.122$ & $0.122$ & $0.140$ & $2.074$ & $2.077$ & $2.194$\\ 
 & TB-mBJ & $3.492$ & $3.485$ & $3.852$ & $0.092$ & $0.091$ & $0.106$ & $1.869$ & $1.867$ & $1.963$\\ 
 \multirow{ 2}{*}{$0.75$} & PBE & $4.045$ & $3.868$ & $4.286$ & $0.113$ & $0.106$ & $0.122$ & $2.011$ & $1.967$ & $2.070$\\ 
 & TB-mBJ & $3.252$ & $3.093$ & $3.371$ & $0.082$ & $0.076$ & $0.087$ & $1.803$ & $1.759$ & $1.836$\\ 
\hline
\hline
\end{tabular}
\end{table}

The real part of the dielectric function ($\epsilon_1(\omega)$), reflectivity ($R(\omega)$) and refractive index ($n(\omega)$) spectrum are shown in FIG. S3 of the supporting information and their static values are listed in the TABLE \ref{tab:e1rnibr}. The average static dielectric constant ($\epsilon_{1}(0)$) calculated with PBE and TB-mBJ are $4.802$, $4.476$, $4.066$; $3.887$, $3.610$ and $3.239$ with the increasing \ch{Br} percentage of $0.25$, $0.50$ and $0.75$, respectively.
The static dielectric constant $\epsilon_{1}(0)$ calculated with PBE of \ch{I}-\ch{Br} mixed \ch{RbPbI_3} systems are larger than that of pristine \ch{RbPbBr_3} but are smaller than that of pristine \ch{RbPbI_3} \cite{ref34}. A large static dielectric constant value will help to reduce the recombination of electron and hole and, thus, will increase the device performance. The average static refractive index ($n(0)$) calculated with PBE and TB-mBJ are $2.191$, $2.115$, $2.016$; $1.971$, $1.900$ and $1.799$ for $\ch{x}=0.25$, $0.50$ and $0.75$, respectively. The refractive index decreases with the increasing bromide percentage for  the considered halide mixed \ch{RbPbI_3} systems. The average static reflectivity ($R(0)$) using both PBE and TB-mBJ are estimated as $13.9$\%, $12.8$\%, $11.4$\%; $10.7$\%, $9.6$\% and $8.2$\% for the increasing \ch{Br} content of $0.25$, $0.50$ and $0.75$, respectively. The reflectivity decreases with the increasing \ch{Br} content which agrees with the fact of the higher reflectivity for \ch{RbPbI_3} than that of \ch{RbPbBr_3} \cite{ref34}.

\begin{table}[h!]
\caption{The calculated average integrated absorption intensity in $10^{3}$ eV/cm for \ch{RbPb(I_{1-x}Br_x)_3}}
\label{tab:iiibr}
\begin{tabular}{ccccccc}
\hline
\hline
& \multicolumn{3}{c}{PBE} & \multicolumn{3}{c}{TB-mBJ}\\
\hline
Energy range (eV) & $0-1.7$ & $1.7-3.3$ & $3.3-5.0$ & $0-1.7$ & $1.7-3.3$ & $3.3-5.0$ \\
\hline
$\ch{x}=0.25$ & $2.420$ & $60.817$ & $1107.416$ & $1.562$ & $14.873$ & $649.619$\\
$\ch{x}=0.50$ & $2.074$ & $38.331$ & $935.044$ & $1.340$ & $9.951$ & $457.450$ \\
$\ch{x}=0.75$ & $1.687$ & $24.472$ & $630.978$ & $1.051$ & $6.525$ & $224.397$ \\
\hline
\hline
\end{tabular}
\end{table}

\begin{table}[h!]
\caption{Effective masses for \ch{RbPb(I_{1-x}Br_{x})_3}}
\label{tab:emibr}
\begin{tabular}{cccccccccc}
\hline
\hline
&  & \multicolumn{2}{c}{\ch{x}=$0.25$} & \multicolumn{2}{c}{\ch{x}=$0.50$} & \multicolumn{2}{c}{\ch{x}=$0.75$} & $E_b$ & $a^*$\\
\hline
& Directions & Electron & Hole & Electron & Hole & Electron & Hole & (meV) & (\AA)\\
\hline
PBE & \multirow{2}{*}{$X \rightarrow S$} & $0.129$ & $0.076$ & $0.347$ & $0.056$ & $0.058$ & $0.158$ & $44.8$ & $33.4$\\
TB-mBJ &  & $0.107$ & $0.289$ & $0.566$ & $0.069$ & $0.117$ & $0.383$ & $58.5$ & $31.6$\\
PBE & \multirow{2}{*}{$Y \rightarrow \Gamma$} & $0.085$ & $0.177$ & $0.346$ & $0.074$ & $0.118$ & $0.127$ & $29.8$ & $53.8$\\
TB-mBJ &  & $0.110$ & $0.292$ & $0.566$ & $0.083$ & $0.118$ & $0.335$ & $56.3$ & $35.3$\\
PBE & \multirow{2}{*}{$\Gamma \rightarrow Z$} & $0.038$ & $2.212$ & $0.068$ & $0.032$ & $0.029$ & $0.252$ & $32.1$ & $55.1$\\
TB-mBJ &  & $0.039$ & $0.256$ & $0.074$ & $0.034$ & $0.034$ & $0.217$ & $147.8$ & $15.0$\\
\hline
\hline
\end{tabular}
\end{table}
 
Moreover, the effective masses of the photo induced electrons and holes are also estimated using both PBE and TB-mBJ potential. The top of the VB and bottom of the CB are fitted to a parabola and then the effective masses are calculated from the following relation:
\begin{eqnarray}
m_{eff} = {\hbar}^{2} \frac{1}{\frac{{\partial}^{2}E(k)}{\partial k^2}}
\end{eqnarray}
where, $k$ and $E(k)$ are the wave vector and the energy corresponding to the top and bottom of the VB and CB, respectively. The calculated effective masses of electrons and holes along the high symmetry k path of $X \rightarrow S$, $Y \rightarrow \Gamma$ and $\Gamma \rightarrow Z$ are listed in TABLE \ref{tab:emibr}. The average effective masses of electrons ($m_e$) and holes ($m_h$) calculated using PBE are $0.084$, $0.254$, $0.068$ and $0.822$, $0.054$, $0.090$ for $\ch{x}=0.25$, $0.50$ and $0.75$, respectively. The reduced effective masses are then estimated using: $\mu=\frac{m_{e} m_{h}}{m_{e}+m_{h}}$ and found to be $0.076$, $0.044$ and $0.039$ for the increasing \ch{Br} percentage as $0.25$, $0.50$ and $0.75$, respectively. Using TB-mBJ potential the average effective masses of electrons and holes are $0.085$, $0.402$, $0.179$ and $0.279$, $0.062$, $0.312$ for $\ch{x}=0.25$, $0.50$ and $0.75$ in the mixed halide systems, respectively. The reduced effective masses resulted from TB-mBJ are $0.065$, $0.54$ and $0.114$ corresponding to $\ch{x}=0.25$, $0.50$ and $0.75$, respectively. The reduced effective mass calculated using PBE decreases with the increasing \ch{Br} content whereas it increases when $\mu$ is calculated from TB-mBJ. The effective masses of \ch{I}-\ch{Br} mixed \ch{RbPbI_3} structures are found to be lesser than that of the pristine \ch{RbPbI_3} \cite{ref34}. \\The effective masses and the static dielectric constants are important parameters to find the binding energy ($E_b$) of the excitons. The binding energy and the radius of the lowest bound state ($a^*$) are estimated using $E_b=\frac{13.6 \times \mu}{\epsilon_{1}(0)^{2}}$ and $a^*=\frac{\epsilon_{1}(0) \times a_0}{\mu}$ (where, $a_0$ = Bohr radius), respectively. All the values of $E_b$ and $a^*$ for all \ch{I}-\ch{Br} mixed systems are listed in TABLE \ref{tab:emibr}. The binding energy of excitons calculated using PBE and TB-mBJ are $44.8$, $29.8$, $32.1$ and $58.5$, $56.3$, $147.8$ for the increasing \ch{Br} percentage as $0.25$, $0.50$ and $0.75$, respectively. The binding energy of excitons are smaller for all the systems except that for $\ch{x}=0.75$ when it is calculated using TB-mBJ potential. The calculated radius of the lowest bound state as listed in TABLE \ref{tab:emibr} are found to be larger than the lattice parameters for all cases except $\ch{x}=0.75$ using TB-mBJ potential. Therefore, the calculations using PBE suggest excitons for all cases to be Mott-Wannier type. The calculations using TB-mBJ predicts the excitons for both $\ch{x}=0.25$ and $0.50$ are of Mott-Wannier type whereas it is of Frenkel type for $\ch{x}=0.75$. This is also agreed well with the fact that excitons using PBE are of Mott-Wannier type for both pristine \ch{RbPbI_3} and \ch{RbPbBr_3} \cite{ref34}.
 
\begin{figure}[h!]
\centering
  \subfloat[SLME estimated with PBE]{\includegraphics[width=6cm,height=7cm]{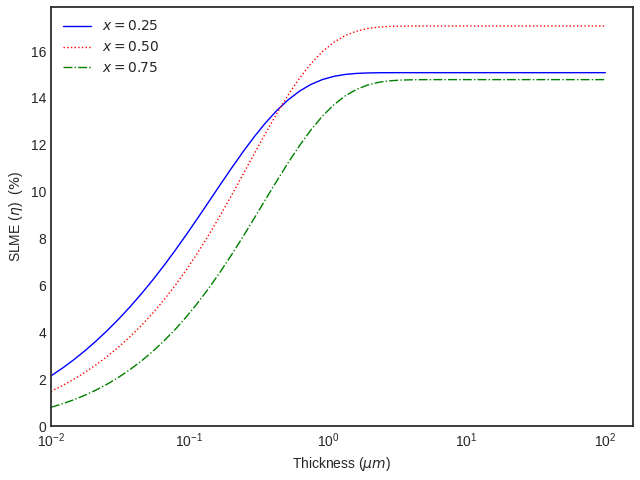}\label{fig:slmepibr}}
  \subfloat[SLME estimated with TB-mBJ]{\includegraphics[width=6cm,height=7cm]{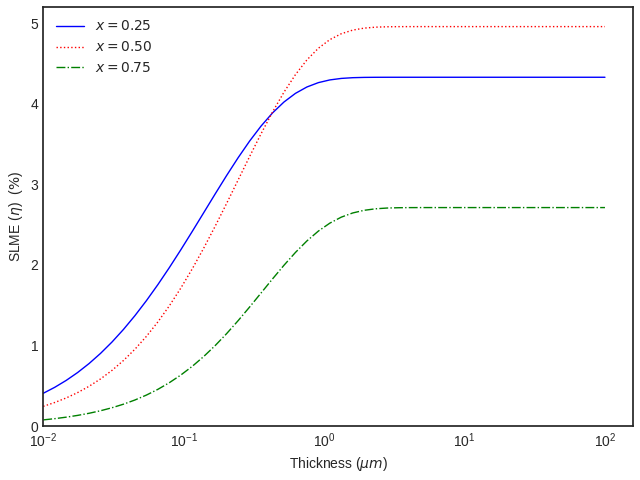}\label{fig:slmemibr}}
     \caption{SLME estimated with PBE and TB-mBJ for \ch{RbPb(I_{1-x}Br_{x})_3}.}\label{fig:slmeibr}
\end{figure}
  
Moreover, the spectroscopic limited maximum efficiency (SLME) are also calculated using both PBE exchange correlation functional and TB-mBJ potential for all the considered halide mixed systems and are shown in FIG. \ref{fig:slmeibr}. The absorption coefficients and bandgaps calculated from the first principle calculations are used to estimate SLME. All the required formulae and input data calculated using the first principle calculations as shown TABLE SII to find SLME are provided in the supporting information. The FIG. \ref{fig:slmeibr} shows that SLME  using both PBE and TB-mBJ at 300 K temperature increases with the increase of the thickness. The estimated SLME at a temperature of 300 K and thickness of 500 nm using PBE and TB-mBJ potential are $13.8$\%, $14.0$\%, $11.1$\% and $3.9$\%, $4.1$\%, $1.96$\% for the increasing \ch{Br} concentrations of $0.25$, $0.50$ and $0.75$, respectively. The equal admixture of \ch{I} and \ch{Br} in \ch{RbPb(I_{1-x}Br_x)_3} shows the highest efficiency (SLME) when it is estimated with both PBE and TB-mBJ. The higher bandgaps estimated using TB-mBJ is responsible for such lowe SLME. 
\section{Conclusion}
In summary, the first principle calculations using PBE and TB-mBJ potential have been carried out to investigate the effect of iodide-bromide mixing in \ch{RbPbI_3} on the electronic and optical properties. The calculated badstructures show all \ch{RbPb(I_{1-x}Br_x)_3} structures have indirect bandgaps. The lowest bandgap of $2.288$ and $2.986$ eV are estimated for $\ch{x}=0.50$ using PBE and TB-mBJ, respectively. PDOS analysis exhibits the ascending contribution of \ch{Br}-4p in the conduction band region with the increasing \ch{Br} concentration. The calculated TDOS depicts a large number of available states for both \ch{RbPb(I_{0.75}Br_{0.25})_3} and \ch{RbPb(I_{0.25}Br_{0.75})_3}, therefore, this enhances the probability of high charge carrier concentrations. The absorption estimated using TB-mBJ also shifts to the higher energy region than that using PBE. The absorption also decreases with increasing \ch{Br} concentration. The effective masses using both PBE and TB-mBJ for all the structures are found to be low and, thus, is beneficial for the ease of the carrier transport. In case of TB-mBJ potential, the binding energy of exciton is high for $\ch{x}=0.75$  whereas it is low for all other cases. The highest SLME of $14.0$\% using PBE and $4.1$\% using TB-mBJ are observed for \ch{RbPB(I_{0.50}Br_{0.50})_3}. \ch{RbPb(I_{0.50}Br_{0.50})_3} shows the lowest bandgap and also highest efficiency than all other structures under our investigation, hence, turns out to be the suitable one for the photovoltaic operation.  

\bibliography{refibr}

\end{document}